# Direct Observation of Localized Radial Oxygen Migration in Functioning Tantalum Oxide Memristors

*Suhas Kumar,[1,2] Catherine E. Graves,[1] John Paul Strachan,[1,*] Emmanuelle Merced Grafals,[1] Arthur L. David Kilcoyne,[2] Tolek Tyliszczak,[2] Johanna Nelson Weker,[3] Yoshio Nishi[4] and R. Stanley Williams[1]*

[1]*Hewlett Packard Laboratories, 1501 Page Mill Rd, Palo Alto, CA 94304, USA*
[2]*Advanced Light Source, Lawrence Berkeley National Laboratory, Berkeley, CA 94720, USA*
[3]*SLAC National Accelerator Laboratory, 2575 Sand Hill Road, Menlo Park, CA 94025, USA*
[4]*Stanford University, Stanford, CA 94305, USA*
*E-mail:* john-paul.strachan@hpe.com

*Oxygen migration in tantalum oxide, a promising next-generation storage material, is studied using in-operando x-ray absorption spectromicroscopy and is used to microphysically describe accelerated evolution of conduction channel and device failure. The resulting ring-like patterns of oxygen concentration are modeled using thermophoretic forces and Fick diffusion, establishing the critical role of temperature-activated oxygen migration that has been under question lately.*

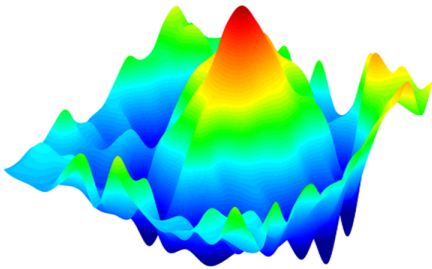

Tantalum oxide memristors are frontrunners for next generation memory technology due to their promise of long endurance, long retention and low power.[1-4] Recent efforts to uncover the nanophysics behind resistance switching in tantalum oxide and several related materials suggest local conductive channel formation by oxygen-ion-migration as the mechanism of operation.[5-9] A more complete understanding of the migration microphysics is required to construct improved compact models that are necessary for circuit design and simulation. In particular, while it is generally acknowledged that elevated local temperatures[10,11] are reached during switching, there is uncertainty on the role of thermally-driven oxygen migration in memristors[12,13] and its influence on eventual device failure.[14,15] Here we present x-ray spectromicroscopy images at the O K-edge and Ta L$_3$-edge of in-operando electrically-cycled TaO$_x$ memristor devices which reveal radial lateral oxygen migration. Following electrical cycling using high-voltage ($\pm$5 V) pulses, we observed irreversible sub-micrometer sized insulating oxygen-rich ring-like features with an oxygen-deficient, highly conductive core. We show that these features are well described by a microphysical model containing thermally-driven diffusion, in the form of thermophoresis and Fick diffusion, during accelerated evolution of the conduction channel. As TaOx devices were further cycled with over 1 million voltage pulses, we observed the oxygen-rich rings breaking into oxygen-rich and oxygen-poor clusters which eventually shorted the electrodes and caused actual device failure. The O K-edge spectra of the oxygen-rich and oxygen-deficient regions in the rings and clusters are consistent with clusters of oxygen interstitials and vacancies, respectively, indicating strong clustering forces. These results directly observe the nanoscale motion of oxygen in a functional memristor especially during the course of accelerated device evolution and failure, and also demonstrate the importance and sign of the thermophoretic forces on oxygen during evolution of conduction channels.

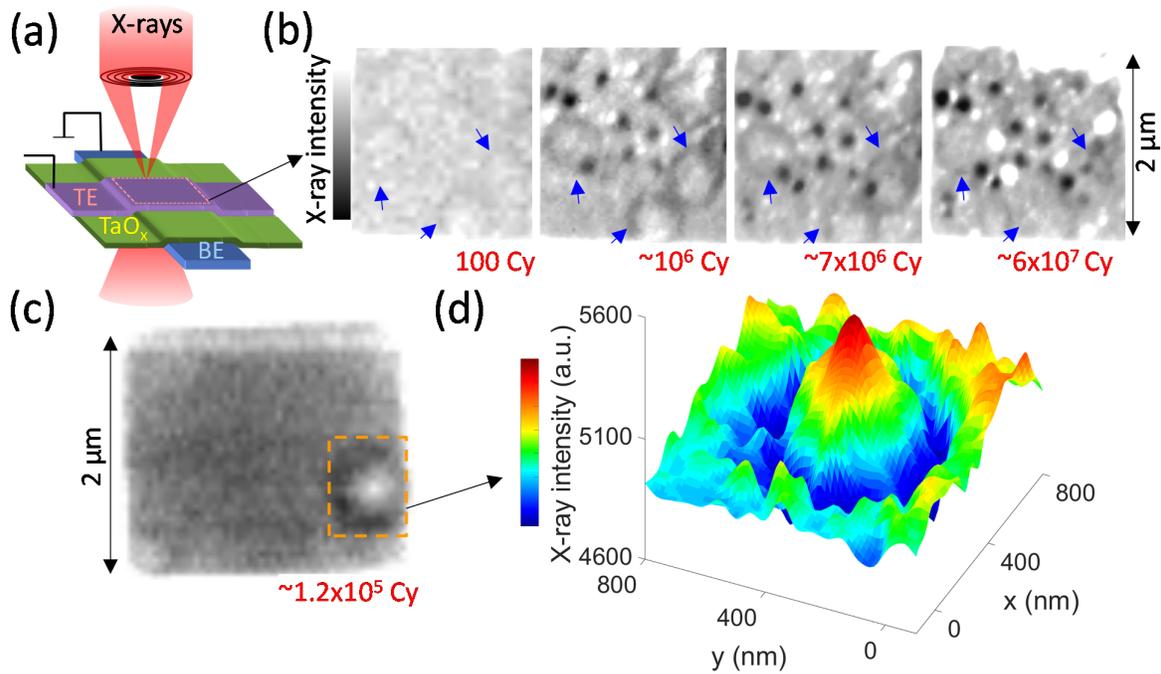

*Figure 1: (a) Schematic of the experimental setup including the x-rays and the crosspoint device. Top electrode (TE), bottom electrode (BE) and tantalum oxide ($TaO_x$) are shown. (b) O K-edge x-ray transmission intensity images of a crosspoint device after different numbers of applied 5 V cycles (Cy). Region of the crosspoint displayed in the data are indicated by a dashed rectangle in (a). All maps were obtained using single x-ray energies between 530 and 533 eV. Blue arrows point to particular ring-like features in the maps. (c) O K-edge transmission intensity map of a different device cycled to 120,000 cycles imaged at energy 531.2 eV. (d) 3-dimensional color-intensity plot of the ring seen in (c) (within the orange dashed rectangle) displaying the profile of the ring.*

The migration of oxygen was investigated using O K-edge x-ray absorption spectromicroscopy[16] on 2 μm x 2 μm crosspoint TaOx memristor devices during in-situ electrical cycling (Figures 1a-b). Memristor devices ($Pt/TaO_x/Ta/Pt$) were specifically fabricated to enable x-ray transmission measurements and exhibited >10 million switching cycles with a $R_{off}/R_{on}$ ratio of 10-100 at low voltage amplitudes (see Supporting Information). To amplify the material changes and accelerate device failure, we used larger amplitude cycling pulses of ±5 V (following device forming and DC switching), and tracked the associated material changes by acquiring x-ray absorption images over the course of several million cycles (Figure 1b). After 100 cycles, x-ray images at the lowest conduction band revealed barely discernible rings with a perimeter darker than the surrounding film area, indicating an increased absorption of the resonantly-tuned x rays in the ring perimeter. As the device reached $10^6$ cycles, dark and bright regions, indicating areas of even greater and reduced x-ray absorption, appeared and grew in number and contrast as the rings faded. We confirmed the ring formation with additional x-ray images on fresh devices after $10^5$ cycles (Figures 1c-1d), before the dark and bright regions appeared, where we again observed a dark ring (increased x-ray absorption) with a bright center (reduced x-ray absorption) relative to the rest of the device. We also further examined the spatial and contrast evolution of the rings and bright and dark regions with increasing electrical cycling (Figures 2a-2b). A prominent dark ring perimeter had appeared by $10^6$ cycles, and then subsequently diminished in intensity by $6×10^7$ cycles (Figure 2b), coinciding with a significant contrast increase of previously existing dark and bright regions and the appearance of a new bright region (at ~0.3 μm). Therefore, the rings appear to fragment to form smaller and more concentrated clustered bright and dark regions. Upon formation of the rings, the device required higher voltages to undergo resistive switching (>5 V) than under normal operation (<2 V). After about $7 × 10^6$ cycles, when a significant number of bright spots were prominently observed, the device became irreversibly stuck in its high conductance state and could not be switched back to a lower conductance state, while the rings themselves were stable over several months. This is an indication that conducting channels observable as bright spots essentially shorted the top and bottom electrodes of the device and caused device failure.

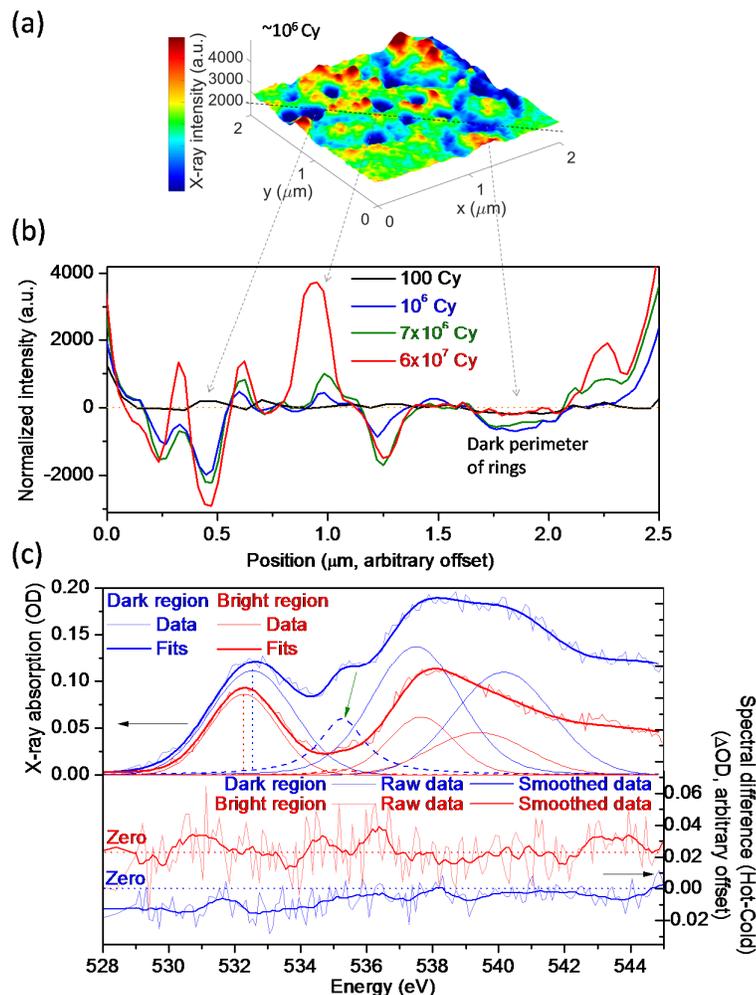

*Figure 2: (a) 3-dimensional color-intensity plot of the O K-edge x-ray image of the device cycled $10^6$ times (from Figure 1b). (b) One dimensional intensity profiles across a particular set of bright and dark spots as a function of voltage cycling, indicated by a dashed black line along which the intensity data was averaged over a width of about 100 nm. Arrows point to some bright and dark spots within the profiled region. Orange dotted line is zero normalized intensity. Also seen is the dark perimeter of two rings that appear within the cross section (~1.75 μm). The absence of a negative peak in the intensity profile (~1.75 μm) at $6×10^7$ cycles shows that the ring disappearance was not an artifact of image contrast stretching in Figure 1b. (c, upper panel) O K-edge absorption spectra (in optical density or OD) collected in the bright and dark regions of the device. The decomposition of the spectra into sub-bands was accomplished by fitting to separate peaks. Vertical dotted lines indicate the position of the lowest conduction band, with corresponding colors. (c, lower panel) Difference between O K-edge spectra with and without current, driven by 5 V pulses (hot state minus cold state), obtained synchronously in the same spatial regions as the spectra shown in the upper panel.*

The bonding and electrical nature of the bright and dark regions of the ring and clusters was investigated with O K-edge absorption spectroscopy (Figure 2c). The spectra were aligned at 528 eV (before the absorption edge) to enable a comparison of changes associated purely with oxygen bonding and oxygen concentration and subtract any electrode effects. The lowest conduction band of the bright region ($\pi^*$, $t_{2g}$), was significantly downshifted (by ~0.3 eV) indicating a higher electrical conductivity than the dark region.[17,18] Also, the bright region spectra revealed lower absorption in the post-edge, indicating a lower oxygen concentration relative to the dark regions. Relative to the rest of the grey crosspoint area, the bright center and dark perimeter of the rings had a -17±2% deficiency and a +14±2% excess of oxygen atoms, and are therefore ascribed to O vacancies and interstitials respectively (see Supporting Information for calculation details). Similarly, the bright and dark regions had a -14±2% deficiency and a +12±2% excess of oxygen atoms (after $6×10^7$ cycles). In a surprising observation, the dark regions displayed a spectral feature in the 535 eV region of the spectrum. This sub-band was significantly smaller, and nearly absent, in the bright region (see the dashed blue sub-band indicated by green arrow in Figure 2c). A similar feature has been previously associated with a superoxide species ($O_2^-$) that binds with an electropositive element such as tantalum,[17,19-24] supporting the existence of oxygen interstitials within the dark regions. Further evidence of this physical picture is indicated by observing where the current flows in the device.

To investigate this, we plot the difference between spectrum with no current flow and spectrum with current flow for the bright and dark regions separately, leveraging an in-operando synchronous time-multiplexed technique developed for this study (Figure 2c).[17] The difference spectra of the bright regions shows more recognizable and sharper features at peak O K-edge energies as compared to that in the dark region, indicating that current preferentially flowed through the bright region, causing a measurable change during joule heating, which is again consistent with reduced oxygen concentration within the bright regions.

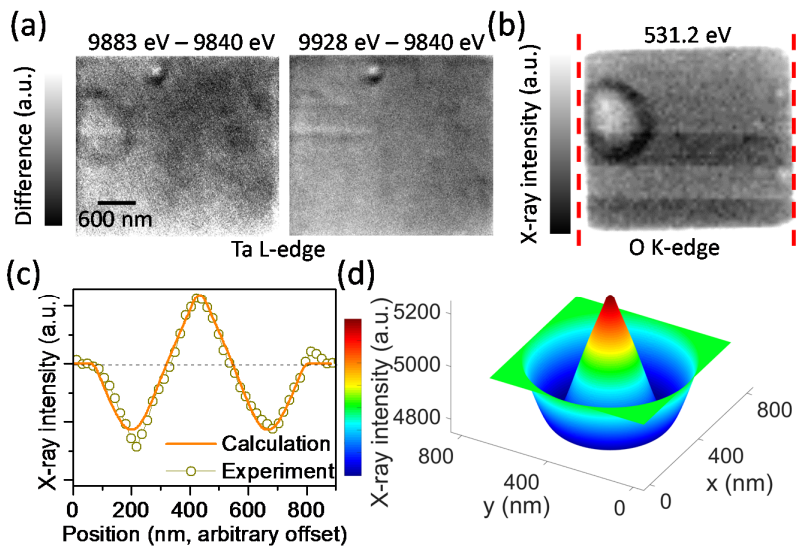

Figure 3: Additional ring formation in a device cycled ~10^5 times (a) Ta L-edge x-ray transmission maps at an energy on the lowest absorption edge (9883.5 eV) and one at a post-absorption edge (9980 eV), both normalized to a pre-Ta absorption energy (9840 eV). The feature near the top edge is an artifact due to beam damage from prior O K-edge measurements (see Supporting Information). (b) O K-edge map of the same device shown in (a). Red dashed lines indicate the edges of the bottom electrode. This panel shows the deformation of the ring pattern correlating with the edge of the bottom electrode. (c) One-dimensional cross section of the ring shown in Figures 1c-1d compared to calculations based on thermophoresis. Dashed horizontal line is the average intensity outside the ring. (d) shows full maps of the ring-formation predicted by calculations to match experimental data of Figure 1d.

The Ta atomic distribution and chemical changes within the rings was investigated utilizing Ta $L_3$-edge spectromicroscopy[25] (Figures 3a) on a fresh device, whose corresponding O K-edge image is shown in Figure 3b. X-ray images were taken at energy points across the Ta $L_3$ absorption edge, revealing an increased x-ray absorption within the oxygen-rich interstitial ring. To separate the increased absorption effects due to increased Ta concentration and oxidation state, we compare the images of on-edge (more sensitive to the Ta oxidation state) and post-edge (more sensitive to Ta elemental concentration), normalized to a corresponding pre-Ta-absorption edge image. The clear dark ring in the on-edge image is consistent with chemical changes to the Ta bonding and oxidation states to accommodate the change in oxygen concentration.[26,27] Additionally, the lack of a ring feature in the post-edge image, indicates that migration of tantalum was less significant relative to migration of oxygen, possibly because of the relatively larger size of Ta (see Supporting Information for more data). As shown in recent investigations,[6,28] it is possible there was Ta migration below the spatial, spectral, and signal resolution of our technique.

The observed oxygen-rich ring can be reproduced by modeling a thermally-driven diffusion of oxygen. A number of forces may act on the oxygen vacancies, for example: electric field-driven vertical drift followed by concentration-gradient-driven lateral diffusion (Fick diffusion),[13] substitutional diffusion[8] and/or thermophoresis.[12,14] To show that a simple combination of two such lateral forces can account for the rings, we use one such example of thermophoresis in a local lateral temperature gradient, such as is expected around a conducting channel, balanced by Fick diffusion. Electric field-driven oxygen migration is usually significant in initiating and sustaining the switching mechanism,[8,29,30] which has been addressed using fully coupled three-dimensional solutions,[12] and is ignored here in light of highlighting a high-power-driven amplified irreversible change to the material, which is essentially a failure mechanism. Our measurements have shown that as-grown films contained numerous incipient oxygen defects, including interstitials and vacancies (Figure S2). Upon application of voltage pulses, as the first percolating current pathway is created there will be a localized temperature profile due to joule heating.[14,31-33] In this model, the resultant lateral temperature gradients produce thermophoretic forces (Soret effect) that attract oxygen vacancies radially inward to the hot conducting channel (against the motion of oxygen interstitials), while Fick diffusion, driven by the resulting gradient of oxygen concentration, balances thermophoretic forces.[8,14,34] We represent this process by a combined continuity equation $\partial n_V/\partial t = \nabla \cdot J_{Fick} + \nabla \cdot J_{Soret}$, where $n_V$ is the vacancy concentration, $J_{Fick}$ and $J_{Soret}$ are fluxes due to Fick diffusion and thermophoresis, respectively. Using this approach, we predict the resultant x-ray intensity profile and compare it to experimental data (Figures 3c-3d), which shows good agreement with each other. Details of the calculation are provided in the Supporting Information based on published analysis of thermophoresis in oxide memristors.[14,35,36] Further experimental support for the thermally-driven nature of the oxygen interstitial rings emerges when we consider that the ring deviates from circular symmetry only at the bottom electrode edge, observed consistently in several devices (highlighted in Figure 3b). Similar features observed in Ref. [37] were attributed to abrupt change in thermal conductance at the bottom electrode edge (see Supporting Information), and suggest that higher temperatures are localized close to the bottom interface (Pt) where resistance switching likely occurs. This observation also asserts the importance of electric-field-driven vertical forces during formation of conducting channel, although in our case, due to irreversible changes and accelerated failure of the device, the continued role of electric-field during application of every voltage pulse diminishes and is hence ignored.

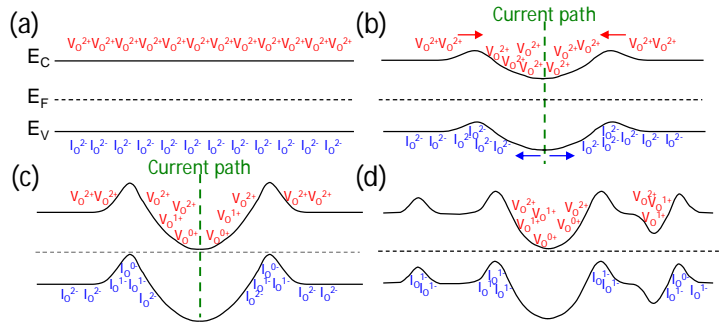

*Figure 4: Schematic illustrations depicting the proposed mechanisms during the formation of rings and dark-bright spots using band diagrams. $V_O$ and $I_O$ represent oxygen vacancies and interstitials, respectively and $E_F$ is the Fermi level. Solid lines represent the conduction ($E_C$) and valence ($E_V$) bands and, for simplicity, the energy levels of $V_O$ and $I_O$ are represented on top of $E_C$ and $E_V$, respectively. (a) Existence of uniformly distributed incipient oxygen interstitials and vacancies. (b) Thermophoretic lateral separation of vacancies and interstitials due to joule heating originating from a filamentary current path. Accumulation of charges creates a change in the local potential energy, causing bending of bands. (c) As the valence and conduction bands approach the Fermi level, some of the charged species are partially or completely neutralized, hence decreasing the repulsion among identical species. (d) Formation of smaller, more stable clusters as the device is cycled.*

In view of device failure, it is important to consider why the like-species of oxygen vacancies and interstitials can cluster together as observed. We propose a schematic picture to enable a better qualitative understanding of the microphysics underlying the observed oxygen migration into rings, and the subsequent formation of clusters of oxygen vacancies and interstitials and the associated band diagram (Figure 4). Initially, incipient oxygen defects are uniformly spread throughout the as-grown $TaO_x$ film (Figure 4a), which then laterally migrate due to thermally-driven forces upon formation of the conducting channel (Figure 4b), as discussed above. Local changes in the electrical potential due to the clustering of negative interstitials and positive vacancies can cause significant bending of the tantalum oxide conduction and valence bands that can partially neutralize the charge on each species,[38] as indirectly evidenced by significant band-shifts in Figure 2c. This band-bending can decrease repulsions among like-charged species and also enable them to agglomerate, as the cohesion energy for oxygen vacancies can be quite low, and is likely to stabilize the ring[7,34,39,40] (Figure 4c). We also calculated an approximate potential profile across the ring (Figure S16) that look qualitatively similar to the one proposed in this cartoon. Subsequent lateral forces followed by clustering of vacancies and interstitials follows the initial bending of bands, as shown in Figures 4b-4c. Due to continued supply of energy through cycling and the large surface-area-to-volume ratio of the rings, they break apart to form clusters of oxygen interstitials and vacancies (Figure 4d), associated with the fading of the ring observed in Figure 2b. Additionally, we observe that most of the bright regions in Figures 1b-1c are in proximity or contact with a dark region, and vice versa (pointed out in Figure S3). This suggests that there are significant vacancy-vacancy and interstitial-interstitial attractive forces or there are significant vacancy-interstitial barriers. The attractive forces likely originate from the strong clustering, as mentioned above. The barrier could originate from the oppositely charged defects in the bright and dark regions behaving as dopants, creating an electric field at the interface of these regions to prevent complete neutralization of the charged defects, much like oppositely charged dopants in a p-n junction. Direct observation of clustering of like-species is an important observation that shines light on a prominent failure mechanism of such devices. We specifically point to the fact that many real world devices are smaller than the ring features observed here and are operated at much lower power levels. Similar experiments utilizing low-power operations on identical devices have yielded strikingly different results, with the device endurance being much higher ($>10^8$) and no rings were observed.[17]

As our measurements show, in-operando x-ray absorption spectromicroscopy is a powerful tool for studying chemical and electronic structure in oxide materials, including device evolution and failure with electrical

Published in Advanced Materials (2016) at http://dx.doi.org/10.1002/adma.201505435

cycling and inhomogeneous localized phenomena. With in-operando, high-voltage electrical cycling of tantalum oxide devices, we observed the development of sub-micrometer features with a ring of oxygen interstitials and an inner core of oxygen vacancies, which could be reproduced using thermally-driven lateral forces. A key observation here is that a significant amount of displaced oxygen moved radially outward from the conduction channel and was stored as interstitials, with a unique spectral signature, in the tantalum oxide film rather than in to the adjacent tantalum metal electrode.[41,42] These results provide experimental data that help in understanding previous models regarding oxygen ion migration,[3,5-8,17,26,32] metastable cohesion of oxygen defects,[7,34,40] role and sign of thermophoresis,[12-14,43] the composition and structure of conduction channels that tend towards failure,[5,6,8,31] and the localization of resistance switching.[14,37,43] Most importantly, we directly observed a failure mechanism caused by clustering of like-species of oxygen.

**Supporting Information**
Supporting Information is available from the Wiley Online Library at
http://onlinelibrary.wiley.com/store/10.1002/adma.201505435/asset/supinfo/adma201505435-sup-0001-S1.pdf?v=1&s=479bb0be81fc9b8a88c310937b2736759bc6674f

**Acknowledgements**
O K-edge spectromicroscopy was performed at Advanced Light Source (ALS) on beamlines 11.0.2 and 5.3.2.2 using focused scanning transmission x-ray microscopes. The ALS is supported by the Director, Office of Science, Office of Basic Energy Sciences, of the U.S. Department of Energy under Contract No. DE-AC02-05CH11231. Ta L-edge spectromicroscopy was performed at Stanford Synchrotron Radiation Lightsource (SSRL) on beamline 6-2 using a transmission x-ray microscope. Use of the SSRL, SLAC National Accelerator Laboratory, is supported by the U.S. Department of Energy, Office of Science, Office of Basic Energy Sciences under Contract No. DE-AC02-76SF00515. The authors gratefully thank Antonio Torrezan, Max Zhang, Kate Norris, Alpha N'Diaye, David Brothers and Zhiyong Li for helping with experimental setup, film growth, x-ray characterization, electron microscopy and useful comments.